\documentclass{article}
\usepackage{verbatim,rotating}
\usepackage{amssymb}
\usepackage{amsmath}

\newtheorem{proposition}{Proposition}
\newtheorem{lemma}{Lemma}
\newtheorem{corollary}{Corollary}

\newcommand{\enproof}{\hfill $\Box$ \vspace*{1ex}}

\newcommand{\ssi}{^{(i)}} 

\newcommand{\mymathbb}[1]{{\mathbb{#1}}} 
\newcommand{\mymathsf}[1]{{\mathsf{#1}}} 

\newcommand{\dmn}{q} 

\newcommand{\sA}{\mymathsf{A}}

\newcommand{\sB}{\mymathsf{B}}

\newcommand{\myF}{{\mymathbb{F}_{\dmn}}} 
\newcommand{\myFnoarg}{{\mymathbb{F}}}

\newcommand{\sS}{\mymathsf{S}}

\newcommand{\cP}{\mathcal{P}}

\newcommand{\sP}{\mymathsf{P}}
\newcommand{\sQ}{\mymathsf{Q}}

\newcommand{\cS}{{\cal S}}
\newcommand{\cT}{{\mathcal T}}

\newcommand{\cX}{{\cal X}}
\newcommand{\sX}{\mymathsf{X}}

\newcommand{\cY}{\myF}  
\newcommand{\cYY}{\myF} 
\newcommand{\cYYY}{{\cal Y}}
\newcommand{\cYpower}[1]{\myFpower{#1}}  
\newcommand{\cW}{{\cal W}}

\newcommand{\vep}{\varepsilon}
\renewcommand{\phi}{\varphi} 
\renewcommand{\subset}{\subseteq}
\renewcommand{\tilde}{\widetilde}

\renewcommand{\bar}{\overline}

\newcommand{\mbm}[1]{\mbox{\boldmath $#1$}} 

\newcommand{\cmple}{^{\rm c}}

\newcommand{\SINT}{\mymathbb{Z}}

\newcommand{\SRN}{\mymathbb{R}}

\newcommand{\Expe}{\mymathsf{E}} 

\newcommand{\Prob}{{\rm Pr}}

\newcommand{\transp}{\mbox{}^{\rm t}}
\newcommand{\supp}{\mymathsf{supp}\,}

\newcommand{\crd}[1]{|#1|}

\newcommand{\dpr}[2]{#1 \cdot #2}

\newcommand{\Acn}[1]{\sA_{#1}}  

\newcommand{\tvara}{s}
\newcommand{\tvarb}{t}

\newcommand{\rvX}{\mymathsf{X}}
\newcommand{\rvY}{\mymathsf{Y}}

\newcommand{\rowspn}{\mymathsf{r}\,}

\newcommand{\rvx}{\sX} 

\newcommand{\vxy}{x}

\newcommand{\rcl}{r_{\rm c}} 

\newcommand{\myFpower}[1]{\mymathbb{F}_{\dmn}^{#1}}

\newcommand{\Qpl}[1]{\sQ^+} 
\newcommand{\Qmi}[1]{\sQ^-}

\newcommand{\Bp}[1]{G}

\newcommand{\zrv}{0_n}

\newcommand{\Bsmall}{\{ 0_n \}} 
\newcommand{\Cgood}{C}
\newcommand{\Jgood}{\tilde{J}}

\newcommand{\crI}{J} 

\newcommand{\Jof}[1]{\prm(\Jgood)}

\newcommand{\tsp}{M}
\newcommand{\tsptwo}[2]{\tsp_{#1}(#2)}

\newcommand{\prm}{\pi}
\newcommand{\rvprm}{\mbm{\pi}}

\newcommand{\cnt}[1]{L_{#1}}
\newcommand{\cpf}{\cP_n}

\newcommand{\Ensperm}{\cS_n}

\newcommand{\Egen}{E} 

\newcommand{\vpp}{P}
\newcommand{\myYpower}[1]{\myFpower{#1}}

\newcommand{\Cite}[1]{\cite{#1}}

\newcommand{\intint}[2]{[#1,#2]\cap{\SINT}} 

\newcommand{\subgrp}{\subset} 

\newcommand{\CSone}{C_1}
\newcommand{\CStwo}{C_2^{\perp}} 
\newcommand{\CStwp}{C_2}

\newcommand{\agd}{A}
\newcommand{\chP}{W}
\newcommand{\kgn}{\kappa}
\newcommand{\Rgn}{r}

\newcommand{\wtgen}{w}

\newcommand{\crW}{\crd{\cW}}

\newcommand{\usedtobetilde}[1]{#1}

\newcommand{\tlJ}{J} 

\title{An Exposition of a Result in\\ ``Conjugate Codes for Secure and Reliable Information
  Transmission''}

\author{Mitsuru Hamada\\[1ex]
Quantum Information Science Research Center,\\
   Tamagawa University Research Institute,\\
6-1-1 Tamagawa-gakuen, 
Machida, Tokyo 194-8610, Japan}

\begin{document}

\maketitle

\begin{abstract}
An elementary proof of the attainability of random coding exponent
with linear codes for additive channels is presented. The result and
proof are from Hamada (Proc.\ ITW, Chendu, China, 2006), and the present
material explains the proof in detail for those unfamiliar with
elementary calculations on probabilities related to linear codes.
\end{abstract}

\section{Introduction}

In this material, 
the details of the proof of a result in \cite{hamada06itw}, 
an article prepared for an invited talk, are
presented without assuming any prerequisite knowledge.
In fact, when the author prepared the manuscript~\cite{hamada07expl},
which includes one illustrative application of the method of
concatenating `conjugate code pairs' devised in \cite{hamada06itw,hamada06md},
the author thought some (or most) proofs are elementary
and straightforward,
so that they are not needed for those working in our society
of information theory.
However, in this article, still more details will be presented
to increase the accessibility.

We remark the result and its detailed proof
are written so that they can be read without referring to \cite{hamada06itw}.
Specifically, in this material,
an elementary proof of the attainability of random coding exponent
with linear codes for additive channels is presented. 
(Of course, many proofs for the attainability of random coding exponent
had existed, 
but the incentive for developing this approach was
to design quantum error-correcting codes and codes that can be used in
cryptographic protocols. For these purposes, we needed to
design codes and decoders under constraints arising from quantum mechanics.)

Thus, this material is supplementary to \cite{hamada06itw}
for those unfamiliar with the elementary approach
adopted in \cite{hamada06itw}, but the result treated in this material
is compact, classical, and comprehensible without understanding the main issues
treated in \cite{hamada06itw}.
This approach is nothing special, but it
may be said to be that of the method of types~\cite{csiszar_koerner,csiszar98},
which requires no prerequisite knowledge,
with the very basics of linear codes incorporated.
%

The aforementioned illustrative application of
the method for concatenation is construction
of pairs of linear codes $(L_1,L_2)$ with $L_2^{\perp} \subset L_1$
(`conjugate code pairs')
that achieve a high information rate on the Shannon theoretic criterion.
Such a code pair can be viewed as a succinct representation of
the corresponding quantum error-correcting code (QECC).
The code construction is explicit in the standard sense that the codes are constructible with polynomial complexity. 
Another (cryptographic) application,
which reflects the original motivation of \cite{hamada06itw,hamada07expl}
has been presented in \cite{hamada08aqa}.

\section{Corrections and Remark to \cite{hamada06itw}} 

\subsection{Corrections to \cite{hamada06itw}; Some Apply Also to \cite{hamada07expl}}

\vspace{1ex}
\begin{enumerate}
\item p.~149, right column, line $14$, `ensemble' should be followed
by `(multiset)'
\item p.~150, left column, line $-1$, 
\[
a_n \crd{\cP_n}^2 d^{-n E_{\rm r}(W,r)}
\]
should read
\[
a_n \crd{\cP_n}^2 q^{-n E_{\rm r}(W,r)}
\]
\item p.~150, right column, line $-9$, `parameter $k$' should read
`the number $k/n$'
\item p.~151, left column, line $-8$, `$(y_1\ssi\cdots y_{N}\ssi)$'
should read `$(y^{(1)} \cdots y^{(N)})$'
\item p.~152, left column, line $1$, 
`$
(\bigoplus_{i=1}^{t} C_1\ssi, \bigoplus_{i=1}^{t} C_2\ssi) 
$'
should read\\
`$
(\bigoplus_{j=1}^{t} C_1\ssi, \bigoplus_{j=1}^{t} C_2\ssi) 
$'
\item p.~152, left column, Eq.~(6),
\[
\tsptwo{Q}{C_j\ssi \setminus \{ 0_n \}} \le (\crd{\cP_n(\myF)}-1) q^{-n(1-r_j)} \agd 
\]
should read
\[
\tsptwo{Q}{C_j\ssi \setminus \{ 0_n \}} \le (\crd{\cP_n(\myF)}-1) q^{-n(1-r_j)} \crd{\cT_{Q}^n} \agd 
\]
\end{enumerate}

Essentially the same errors as in 1, 2 and 6 exist in 
Section~4 of \cite{hamada07expl} (ver.~2),
but the contents of Section~4 of \cite{hamada07expl} 
are presented below in the corrected form.

\subsection{Remark to \cite{hamada06itw,hamada07expl}}

Note that, in \cite{hamada06itw,hamada07expl}, an ensemble has been represented as a multiset, which
is similar to a usual set but permits duplicated entries.

Now the author thinks representing an ensemble as an ordered set
is more natural, as will be done in the present article.

\section{Preliminaries\label{ss:pre}}

In this section, we fix our notation, and recall some notions to be used.
As usual, $\lfloor a \rfloor$ denotes the largest
integer $a'$ with $a'\le a$, and $\lceil a \rceil = - \lfloor - a \rfloor$.
An $[n,k]$ linear (error-correcting) code over a finite field $\myF$, the finite field of $\dmn$
elements, is a $k$-dimensional subspace of $\myFpower{n}$.
The dual of a linear code $C \subset \myFpower{n}$ is
$\{ y \in\myFpower{n} \mid \forall x\in C, \ \dpr{x}{y}=0 \}$
and denoted by $C^{\perp}$,
where $x \cdot y=x y\transp$ with $y\transp$ being the transpose
of $y$.
The zero vector in $\myFpower{n}$ is denoted by $\zrv$. 
The $n \times n$ identity (resp.\ zero) matrix is denoted by $I_n$ (resp.\ $O_n$).
For integers $i \le j$, we often use the set
$\intint{i}{j}=\{ i,i+1,\ldots,j \}$,
which consists of integers lying in the interval $[a,b]
= \{ z \in \SRN \mid a \le z \le b \}$.

We denote the type of $x\in\myFpower{n}$ by $\sP_x$~\cite{csiszar_koerner,csiszar98}.
This means that the number of appearances of $u\in\myF$ in $x\in\myFpower{n}$ is $n\sP_x(u)$.
The set of all types of sequences in $\cYpower{n}$
is denoted by $\cP_n(\cY)$.
Given a set $C\subset\cYpower{n}$, we put 
$\tsptwo{Q}{C}= \crd{\{ y\in C \mid \sP_y = Q \}}$
for types $Q\in\cP_n(\cY)$. 
The list of numbers $(\tsptwo{Q}{C})_{Q\in\cP_n(\cY)}$ 
may be called the $\sP$-spectrum (or simply, spectrum) of $C$.
For a 
type $Q$,
we put $\cT_{Q}^n=\{ y \in\cYpower{n} \mid \sP_{y}=Q \}$. 
We denote by $\cP(\cYYY)$ the set of all probability distributions
on a set $\cYYY$. 
The entropy of a probability distribution $P$ on $\cYYY$ is denoted by $H(P)$,
viz., $H(P)= \sum_{y\in\cYYY} -P(y) \log P(y)$.
Throughout, logarithms are to base $q$.

We follow the convention to
denote by $P_{\rvX}$ the probability distribution of a random variable $\rvX$.

\section{Good Codes in a Balanced Ensemble\label{ss:app_good_spectrum}}

\subsection{Balanced Ensemble}

We can find good codes in an ensemble if the ensemble is `balanced'
in the following sense. 
Suppose $\sS= \{ C\ssi \}_{i=1}^N$ 
is an ensemble (ordered set) of subsets of $\myFpower{n}$.
If there exists a constant $V$ such that
$\crd{\{ i \in \intint{1}{N} \mid x \in  C\ssi \}}=V$ for
any word $x \in \myFpower{n}\setminus\{ \zrv \}$,
the ensemble $\sS$ is said to be {\em balanced}.
(We remark that 
the `balancedness' is defined in a different manner
in \cite{DelsartePiret82} for ensembles of encoders,
not codes.)

The first task in \cite{hamada06itw} was to construct
a relatively small balanced ensemble.
This result can be found in \cite{hamada06itw,hamada07expl},
but it is included in Appendix~\ref{ss:EF}.
With the method of types, we will show
that a large portion of a balanced ensemble consists of good codes.
While the goodness of codes should be evaluated by the decoding error
probability, it is also desirable to quantify the goodness
in such a way that the goodness does not depend on characteristics
of channels. In view of this, the following proposition is useful.

The next proposition relates the spectrum of a code
with its decoding error probability when it is used
on an additive memoryless channel.

\begin{proposition} {\rm \cite[Theorem~4]{hamada05qc}}. \label{th:rcex}
Suppose we have an $[n,\kgn]$ linear code $C$ over $\myF$ such that 
\[
\tsptwo{Q}{C}
\le a_n \dmn^{\kgn-n} \crd{\cT_{Q}^n},
 \quad Q\in\cP_n(\cYY)\setminus \{ \sP_{0_n} \}
\]
for some $a_n \ge 1$. Then, 
its decoding error probability with the minimum entropy syndrome decoding
is upper-bounded by 
\[
a_n \crd{\cP_n(\cYY)}^2 q^{-n E_{\rm r}(\chP,\Rgn)}
\]
for any additive channel $\chP$ of input-output alphabet\/ $\myF$,
where $\Rgn=\kgn/n$ and $E_{\rm r}(\chP,\Rgn)$ is the random coding exponent of 
$\chP$
defined by
\[
E_{\rm r}(\chP,\Rgn) = \min_{Q\in\cP(\cY)} [D(Q||\chP) + |1-\Rgn-H(Q)|^+].
\]
Here, $D$ and $H$ denote the relative entropy and entropy, respectively,
and $|x|^+ =\max\{ 0, x \}$.
\end{proposition}

For a poof, see Section~\ref{app:rcex}.
In the simplest case where $\dmn=2$, the premise of the above proposition
reads `the spectrum
of $C$ is approximated by the binomial coefficients $\crd{\cT_{Q}^n}$ 
up to normalization.'

The following lemma shows a large portion of a balanced 
ensemble $\{ C\ssi \}_{i=1}^{N^*}$ is made of good codes
(we have applied this fact to ensembles written as $\{ C_j\ssi \}_{i=1}^{N^*}$ in \cite{hamada06itw,hamada07expl}).
\begin{lemma}\label{lem:est_bad} \cite[p.~152, left column]{hamada06itw}.
Assume we have a balanced ensemble $\{ C\ssi \}_{i=1}^{N^*}$.
Let us say an $[n,\kappa]$ code
$C\ssi$ is $\agd$-good if 
\begin{equation}\label{eq:a-good0}
\tsptwo{Q}{C\ssi } 
\le \agd (\crd{\cP_n(\myF)}-1) q^{-n(1-\rho)}\crd{\cT_{Q}^n}
\end{equation}
for all $Q\in\cP_n(\myF)\setminus \{ \sP_{0_n} \}$, where $\rho = \kappa/n$.
Then, the number of codes that are not $\dmn^{\vep n}$-good
in $\{ C\ssi \}_{i=1}^{N^*}$ is at most 
\begin{equation}\label{eq:frac_bad0}
z=\lfloor N^* \dmn^{-\vep n} \rfloor.
\end{equation}
\end{lemma}

This lemma will be proved in Section~\ref{th:routineITproofs}.
Note, owing to Proposition~\ref{th:rcex},
for the $\dmn^{\vep n}$-good codes $C\ssi$ in the above lemma, 
the decoding error probability
is upper-bounded by
\begin{equation}\label{eq:bound_inner}
a'_n \dmn^{-n[E_{\rm r}(\chP,\rho)-\vep]}, 
\end{equation}
where $a'_n = \crd{\cP_n(\myF)}^3$ 
is at most polynomial in $n$.

\subsection{Proof of Lemma~\ref{lem:est_bad}  \label{th:routineITproofs}}

A proof of Lemma~\ref{lem:est_bad}
will be given, though it may be a routine in information theory.
%
We have a lemma.

\begin{lemma}\label{prop:ens1u}
Assume $\sS$ and $\cW$ are finite sets, and non-negative 
numbers $f_{\wtgen}(x)$ 
are associate with each pair $(x,\wtgen) \in \sS \times \cW$.
Denote by $\bar{f}_\wtgen$ the average of $f_\wtgen(x)$ over $\sS$:
\[
\bar{f}_\wtgen = \frac{1}{\crd{\sS}} \sum_{x\in\sS} f_\wtgen(x).
\]
Then, for any $a>0$, 
the number of members in $\sS$ that fail to satisfy the condition
\[
 \forall \wtgen\in \cW, \quad f_\wtgen(x) \le \bar{f}_\wtgen \crW a
\]
is upper-bounded by $a^{-1}\crd{\sS}$.
\end{lemma}

{\em Proof.}\/
Let $X$ be a random variable uniformly distributed over $\sS$.
Then, the probability that $X$ fails to satisfy
`$\forall \wtgen\in \cW, f_\wtgen(X) \le \bar{f}_\wtgen \crW a$'
is upper-bounded as follows:
\begin{eqnarray}
\lefteqn{\!\!\!\!\! \!\!\!\!\! \!\!\!\!\! 
\Prob \{ \exists w \in \cW, \, f_\wtgen(X) > \bar{f}_\wtgen \crW a \} }
\nonumber\\
&\le & \sum_{w} \Prob \{ f_\wtgen(X) > \crW \bar{f}_\wtgen a \}  \nonumber\\
&\stackrel{(i)}{=} & \sum_{w:\, \bar{f}_\wtgen >0 } \Prob \{ f_\wtgen(X) > \crW \bar{f}_\wtgen a \}\nonumber\\
&\stackrel{(ii)}{\le} & \sum_{w:\, \bar{f}_\wtgen >0 } (\crW a)^{-1} \le a^{-1} , \label{eq:ens1}
\end{eqnarray}
where the equality $(i)$ and inequality $(ii)$ follow from the
fact that $\bar{f}_\wtgen =0$ implies 
$f_\wtgen(x) = \bar{f}_\wtgen \crW a=0$ for all $x\in \sS$,
and Markov's inequality, respectively. 
Markov's inequality is included at the end of this subsection with a proof.
The lemma immediately follows from (\ref{eq:ens1}).
\enproof

{\em Proof of Lemma~\ref{lem:est_bad}.}\/
From the fact that $\{ C\ssi \}_{i=1}^{N^*}$ is balanced, it follows
\begin{equation}\label{eq:av}
\frac{1}{N^*} \sum_{i=1}^{N^*} 
M_Q(C\ssi) = \frac{q^{\kappa}-1}{q^n-1} \crd{\cT_Q^n}
\le \frac{q^{\kappa}}{q^n} \crd{\cT_Q^n}
\end{equation}
for any $Q\in\cP_n(\myF)$, $Q \ne \sP_{\zrv}$.
To see this, let $V$ be the number of appearances of any fixed nonzero word
in enumerating codewords in $C\ssi$, $i\in \intint{1}{N^*}$.
Then, we have trivial equalities
$V (q^n-1) = N^* (q^{\kappa}-1)$ and
\[
\sum_{i=1}^{N^*} M_{Q}(C\ssi) = V \crd{\cT_{Q}^n}
\]
for any $Q\in\cP_n(\myF)$, $Q \ne \sP_{\zrv}$.%
\footnote{The relation $V (q^n-1) = N^* (q^{\kappa}-1)$ immediately follows
by counting the pairs $(x,C)$ such that $x\in C\setminus \{ 0_n \}$ and $C$ is
a component of $\{ C\ssi \}_{i=1}^{N^*}$ in two ways, and the other equality follows similarly.}
From these, we readily obtain the equality and hence the inequality in (\ref{eq:av}).
Now Lemma~\ref{lem:est_bad} follows
upon applying Lemma~\ref{prop:ens1u} to 
$\sS=\{ (C\ssi, i) \mid i\in\intint{1}{N^*}\}$,
where
$f_\wtgen((C,i))= \tsptwo{Q}{C}$, $\wtgen=Q$
and $\cW=\cP_n(\myF)\setminus \{ \sP_{\zrv} \}$.
\enproof

\begin{lemma}[Markov's Inequality]  \label{lem:Markov}
For a positive constant $A$, 
and a random variable $\rvY$ that takes non-negative values and has
a positive mean $\mu$, we have 
\[
\Prob\{ \rvY \ge A \mu \} \le 1/A .
\]
\end{lemma}

{\em Proof}.\/
We have
$\mu = \sum_{w} P_{\rvY} (y) y \ge \sum_{y:\, y \ge \mu A} P_{\rvY} (y) y
\ge \sum_{y:\, y \ge \mu A} P_{\rvY} (y) \mu A$ 
$=  \mu A \sum_{y:\, y \ge \mu A} P_{\rvY} (y)
=  \mu A \, \Prob \{ \rvY \ge A \mu \}$, which implies the lemma.
\enproof

\subsection{Proof of Proposition~\ref{th:rcex} \label{app:rcex}}

We use the following basic inequality~\cite{csiszar_koerner,csiszar98,cover_th}:\begin{equation}\label{eq:prob_type}
\sum_{y\in\cYpower{n}:\, \sP_y=Q} \vpp^{n}(y) \le \dmn^{-nD(Q||\vpp)}
\end{equation}
for any $P \in \cP(\cY)$.
(Recall $P^n$ denotes the product of $n$ copies of $P$.)
The symmetric group on $\{ 1,\dots, n\}$, 
which is composed of all permutations on $\{ 1,\dots, n\}$,
is denoted by $\cS_n$.
We define an action of $\cS_n$ on $\cYpower{n}$
by
\[
\prm((x_1,\dots,x_n))=(x_{\prm(1)},\dots,x_{\prm(n)})
\]
for any $\pi\in\cS_n$ and $(x_1,\dots,x_n)\in\cYpower{n}$, and put
\[
\prm(C)=\{ \prm(x) \mid x\in C \},
\quad \prm\in\cS_n, \, C\subset\cYpower{n}.
\]
The expectation operation with respect to a random variable $\rvx$
taking values in $\cX$
is denoted by $\Expe_{\rvx}$:
\[
\Expe_{\rvx} f(\rvx) = \sum_{x \in \cX} P_{\rvx}(x) f(x)
\]
where $f$ is a real-valued function on $\cX$.

\begin{lemma}\label{lem:gen}
Assume a linear code $\Cgood \subgrp \cYpower{n}$ 
satisfies
\[
\tsptwo{Q}{\Cgood\setminus \{ 0_n \}}/\crd{\cT_{Q}^n} 
\le a_n \dmn^{-n T},
\quad \quad Q\in\cP_n(\cY)
\]
with some real numbers 
$a_n \ge 1$ and $T$.
Let $\crI$ be a set of coset representatives 
for $\cYpower{n}/\Cgood$ such that
each coset $D\in\cYpower{n}/\Cgood$ has
a representative that belongs to $\crI$ and that attains the minimum of 
$H(\sP_{x})$, $ x\in D$
(the resulting decoding is called minimum entropy decoding).
Then,
we have for any $P_n\in\cP(\cYpower{n})$,
\[
\Expe_{\rvprm} P_n(\rvprm(\crI)\cmple)
\le
a_n |\cP_n(\cY)|
\sum_{Q\in\cP_n(\cY)} P_n(\cT_Q^n) \dmn^{-n |T-H(Q)|^+}
\]
where ${\rm c}$ denotes complement,
$|t|^+ =\max\{t,0\}$, and the random variable $\rvprm$
is uniformly distributed over $\cS_n$.
\end{lemma}

\begin{corollary}\label{coro:gen2}
Assume for a linear code $\Cgood \subgrp \cYpower{n}$, 
$\tsptwo{Q}{\Cgood\setminus\{ 0_n \}}$ is bounded as in Lemma~\ref{lem:gen}.
Then, 
with $\crI$ 
as in the lemma,
we have for any $P\in\cP(\cY)$,
\[
P^n(\crI\cmple) 
\le 
a_n |\cP_n(\cY)|^2 \dmn^{-n \Egen(P,T)}
\]
where 
\[
\Egen(P,T) = \min_{Q\in\cP_n(\cY)} [D(Q||P) + |T-H(Q)|^+].
\]
\end{corollary}

A proof of Lemma~\ref{lem:gen} is given in the next subsection.

{\em Proof of Corollary~\ref{coro:gen2}}.\/
Clearly, $\Expe_{\rvprm} P^n(\rvprm(\usedtobetilde{\crI})\cmple)$
$=$ $P^n(\usedtobetilde{\crI}\cmple)$. 
Then,  inserting the estimate of $P^n(\cT_Q^n)$ 
in (\ref{eq:prob_type}) into the bound on 
$\Expe_{\rvprm} P^n(\rvprm(\usedtobetilde{\crI})\cmple)$ in 
the lemma, we have
\[
P^n(\usedtobetilde{\crI}\cmple) \le a_n \crd{\cP_n(\cY)}
\sum_{Q \in \cP_n(\cY)}{\dmn}^{ - n [D(Q||\vpp)+|T-H(Q)|^+ ] } 
\]
and hence, the corollary.

Putting $T=1-\kappa/n$ in this corollary, we readily obtain the proposition.

\subsection{Proof of Lemma~\protect\ref{lem:gen}}

In the proof, $\cP_n(\cY)$ is abbreviated as $\cpf$.
We will show that 
$\Bp{\vpp}=
\Expe_{\rvprm} P_n(\rvprm(\tlJ)\cmple)$ 
is bounded 
above by the claimed quantity.

Imagine we list up all words in $\prm(\Cgood\setminus\Bsmall)$ for all $\prm\in\Ensperm$
permitting duplication.
Clearly,
the number of appearances of any fixed word $y\in\cYpower{n}$ in the list 
only depends on its type $\sP_{y}\in\cpf$.
Namely, for any $Q\in\cpf$, 
there exists a constant, say $\cnt{Q}$, such that
\begin{equation}\label{eq:a}
\crd{\{\prm\in\Ensperm \mid y\in \prm(\Cgood\setminus\Bsmall) \}}=\cnt{Q}
\end{equation}
for any word $y$ with $\sP_y=Q$. 
Then, counting the number of words of a fixed type $Q$ in the list in
two ways,
we have 
$\crd{\cT_{Q}^n} \cnt{Q} = \crd{\Ensperm} \tsptwo{Q}{\Cgood\setminus\Bsmall}$.
Hence, for any type $Q \in \cP_n(\cY)$
\begin{equation}\label{eq:b}
\frac{\cnt{Q}}{\crd{\Ensperm}}=\frac{\tsptwo{Q}{\Cgood\setminus\Bsmall}}{\crd{\cT_{Q}^n}}
\le a_n \dmn^{-nT} 
\end{equation}
by assumption.
From (\ref{eq:a}) and (\ref{eq:b}),
we have
\begin{equation}\label{eq:balanced_perm}
\frac{\crd{\Acn{y}(\Cgood\setminus\Bsmall)}}{\crd{\Ensperm}} \le a_n \dmn^{-nT}
\end{equation}
for any $y\in\myYpower{n}$, 
where
\[
\Acn{y}(\Cgood\setminus\Bsmall) = \big\{ \prm \in \Ensperm \mid y \in \prm(\Cgood\setminus\Bsmall) \big\}.
\]

Then, we have
\begin{eqnarray}
\Bp{\vpp} && = \frac{1}{\crd{\Ensperm}} \sum_{\prm\in\Ensperm} \sum_{\vxy \notin \tlJ}  P_n(\vxy) \nonumber\\
     & & = \sum_{x \in \myYpower{n}} P_n(x) \frac{ \crd{\{ \prm \in
     \Ensperm \mid
     x \notin \tlJ \}} }{\crd{\Ensperm}}.\label{eq:pr0}
\end{eqnarray}
Since $x \notin \tlJ$ occurs only if there
 exists
a word $u\in\myYpower{n}$ such that $H(\sP_u) \le H(\sP_x)$ and $u-x\in
\prm(\Cgood \setminus \Bsmall)$
from the design of $\tlJ$
specified above (minimum entropy decoding),
it follows
\begin{eqnarray}
\lefteqn{ \!\!\! \crd{ \{ \prm \in \Ensperm \mid x \notin \tlJ \}}/\crd{\Ensperm} }\nonumber\\
 &\le &\sum_{u\in \myYpower{n} :\, H(\sP_u) \le  H(\sP_x)}
 \crd{\Acn{u-x}(\Cgood\setminus\Bsmall)}/\crd{\Ensperm}\nonumber\\
  &\le & \sum_{u\in \myYpower{n} :\, H(\sP_u) \le  H(\sP_x)} 
a_n \dmn^{-nT} \nonumber\\
 & = & \sum_{Q'\in \cpf:\, H(Q') \le  H(\sP_x)}
a_n \crd{\cT_{Q'}^n}\dmn^{-nT} \nonumber\\
 & \le & \sum_{Q'\in \cpf:\, H(Q') \le  H(\sP_x)}
a_n \dmn^{nH(Q')-nT} 
\label{eq:pr2} 
\end{eqnarray}
where we have used 
(\ref{eq:balanced_perm}) for the second
inequality, and another well-known inequality~\cite{csiszar_koerner,csiszar98,cover_th}
\begin{equation}\label{eq:typesC}
\forall Q\in \cP_n(\cY),\quad
|\cT_{Q}^n| \le \dmn^{nH(Q)}
\end{equation}
for the last inequality.
Then, 
using the inequalities
$\min \{ a \tvarb, 1 \} \le a \min \{ \tvarb, 1 \}$ and
$\min \{ \tvara+\tvarb, 1\} \le \min \{ \tvara, 1\} + \min \{ \tvarb, 1\}$ for $a \ge 1, \tvara,\tvarb \ge 0$,
we can proceed from (\ref{eq:pr0}) as follows, which completes the proof:
\begin{eqnarray*}
\Bp{\vpp}&\le& \sum_{x\in\myYpower{n}} P_n(x) \min \Bigl\{
\sum_{ Q'\in \cpf:\, H(Q') \le
H(\sP_x) } 
a_n \dmn^{nH(Q')-nT } 
,\ 1 \ \Bigr\}\\
 &\le &  a_n \sum_{Q\in\cpf} P_n(\cT_Q^n) 
\min \Bigl\{
\sum_{Q'\in \cpf:\, H(Q') \le
H(Q)} \!\!\!\!\!
\dmn^{nH(Q')-nT},
1 \Bigr\}\\
 &\le &  a_n \sum_{Q\in\cpf} P_n(\cT_Q^n) 
\sum_{Q'\in \cpf:\, H(Q') \le H(Q)} 
\min \bigl\{ \dmn^{-n[T-H(Q')]}  
,\ 1 \ \bigr\}\\
&\le & a_n \crd{\cpf} \sum_{Q\in\cpf} P_n(\cT_Q^n) 
\max_{Q'\in \cP(\myF):\, H(Q') \le
H(Q) } {\dmn}^{- n |T-H(Q')|^+ }\\
&=&  a_n \crd{\cpf}  \sum_{Q\in\cpf} P_n(\cT_Q^n) 
{\dmn}^{-  n |T-H(Q) |^+ }.
\end{eqnarray*}

\section{Concluding Remarks \label{ss:sum_rem}} 

In \cite{hamada06itw,hamada06md} (or \cite{hamada07expl}), 
quantum-mechanically compatible pairs of linear codes that are constructible
with polynomial complexity were presented. 
The Calderbank-Shor-Steane quantum codes corresponding to the constructed pairs achieve
the so-called Shannon rate.
The most novel result among these would be
the method for concatenating compatible (conjugate) code pairs, 
which have been published in \cite{hamada06md}. 

The present material was prepared
for explaining the results not included in \cite{hamada06md} 
for those unfamiliar with the elementary combinatorial approach
(the method of types with the very basics
of linear codes incorporated).

This material 
might be included somewhere else (possibly in some other context).

\appendix

\section{Some Other Contents of \cite{hamada06itw}}

\subsection{Compatible (Conjugate) Code Pairs \cite{hamada06itw}}

Consider a pair of linear codes $(C_1,C_2)$ satisfying
\begin{equation}\label{eq:css_cond}
\CStwo \subgrp \CSone, 
\end{equation}
which condition is equivalent to $\CSone^{\perp} \subgrp \CStwp$.
The following question arises from
an issue on quantum error correction:
How good
both $C_1$ and $C_2$ can be under the constraint
(\ref{eq:css_cond})? This is the subject treated in \cite{hamada06itw,hamada06md,hamada07expl}.

We have named a pair $(C_1,C_2)$ with (\ref{eq:css_cond})
a conjugate code pair in \cite{hamada06itw}.
In what follows,
we will use a `compatible code pair' in place of `conjugate code pair.'

\subsection{Code Ensemble Based on Extension Field~\cite{hamada06itw} \label{ss:EF}}

The companion matrix of a polynomial
$f(x)=x^{n}-f_{n-1}x^{n-1}-\cdots-f_1 x -f_0$,
which is monic (i.e., of which the leading term has coefficient 1), 
over $\myF$
is defined to be
\[
T=\begin{bmatrix} 
&0_{n-1} & f_0 \\
&{\Large I_{n-1}} & \begin{matrix} f_1 \\ \vdots \\ f_{n-1} \end{matrix}
\end{bmatrix}.
\]
Let $T$ be the companion matrix, or its transpose,
of a monic primitive 
polynomial of degree $n$ over $\myF$.
Given an $n\times n$ matrix $M$, let $M|^{m}$ (resp.\ $M|_m$) denote the $m\times n$ submatrix
of $M$ that consists of the first (resp.\ last) $m$ rows of $M$.
We put $C_1^{(i)}=\{ x T^i|^{k_1} \mid x\in\myFpower{k_1} \}$ and
$C_2^{(i)}=\{ x (T^{-i})\transp|_{k_2} \mid x\in\myFpower{k_2} \}$
for $i=1,2,\dots$,
where $M\transp$ denotes the transpose of $M$.
Then, setting
\begin{equation}\label{eq:ensT}
\sB=\sB_T=\{ (C_1^{(i)},C_2^{(i)}) \}_{i=1}^{\dmn^n-1} ,
\end{equation}
we have the next lemma. 

\begin{lemma}{\rm \cite[Lemma~1]{hamada06itw}}. \label{lem:ens0}
Let $T$ be the companion matrix
of a monic primitive polynomial of degree $n$ over $\myF$.
For 
integers $k_1,k_2$ with $0 \le n-k_2 \le k_1 \le n$
and $\sB_T=\{ (C_1\ssi,C_2\ssi) \}_{i=1}^{\dmn^n-1}$ constructed as above, 
any $(C_1\ssi,C_2\ssi)$ is
a compatible code pair, and 
both $\{ C_1\ssi \}_{i=1}^{\dmn^n-1}$
and $\{ C_2\ssi \}_{i=1}^{\dmn^n-1}$ are balanced.
\end{lemma}

{\em Remark.}\/
It is known (and proved in a self-contained manner in \cite[Sections~VII]{hamada06md}) that the matrix $T$ 
has the following property, 
which are used in the proof of Lemma~\ref{lem:ens0} below:
The set $\{ O_{n}, I_{n}, T, \ldots, T^{\dmn^{n} -2} \}$ 
is isomorphic to $\myFnoarg_{\dmn^n}$ as a field. \enproof

\begin{figure}[t]
\begin{center}
\mbox{}\hspace*{2cm}
\scalebox{0.8}{\includegraphics
{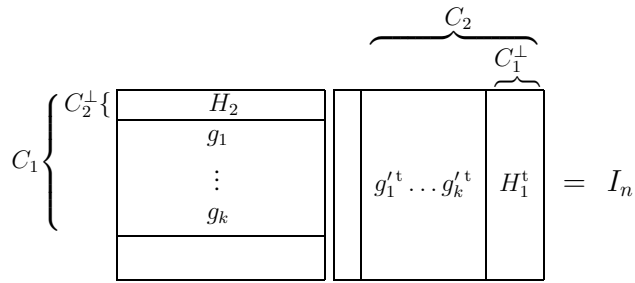} 
}
\caption{A basic structure of an $[[n,k]]$ compatible code pair.
 \label{fig:bs_ccp}\label{fig:css}}
\end{center}
\end{figure}

{\em Proof of Lemma~\ref{lem:ens0}}\/~\cite{hamada06itw}.
The condition (\ref{eq:css_cond}) is fulfilled
since $T^i T^{-i}=I_n$ implies that
the $C_2\ssi\mbox{}^{\perp}$ is spanned by the first $n-k_2$ rows of $T^i$.
(This is easily seen if we divide the two matrices 
on the left-hand side of $T^i T^{-i}=I_n$
into submatrices as in Figure~\ref{fig:bs_ccp}.)

We can write 
$C_1^{(i)}=\{ y T^i \mid y\in\myFpower{n},\, 
\supp y \subset \intint{1}{k_1} \}$,
where $\supp (y_1,\dots,y_n)$ $=$ $\{ i \mid y_i \ne 0 \}$. 
Imagine we list up all codewords in $C_1^{(i)}$ permitting duplication.
Specifically, we list up all $y T^i$
as $y$ and $i$ vary over the range
$\{ y  \mid y\in\myFpower{n},\, \supp y \subset \intint{1}{k_1} \}$
and over $\intint{1}{\dmn^n-1}$, respectively.

With $y\in\myFpower{n}\setminus \{ 0 \}$ fixed, 
$y T^i$, $i\in\intint{1}{\dmn^{n}-1}$, are all distinct
since $T^i \ne T^j$ implies $y T^i - y T^{j}=y T^{l}$ for some 
$l$ and $y T^{l}$ is not zero.
Hence, any nonzero fixed word in $\myFpower{n}$ appears
exactly $\dmn^{k_1}-1$ times in listing $yT^i$ as above.
Namely, the ensemble $\{ C_1\ssi \}_{i=1}^{\dmn^n-1}$ is balanced.
Using $(T^{-i})\transp$ in place of $T^i$, we see
the ensemble $\{ C_2\ssi \}_{i=1}^{\dmn^n-1}$ is also balanced, completing the proof.
\enproof

Lemmas~\ref{lem:est_bad} and \ref{lem:ens0} show
the existence of a compatible code pair having exponentially decreasing decoding error
probabilities in $\sB$.


\begin{thebibliography}{1}

\bibitem{hamada06itw}
M.~Hamada, ``Conjugate codes for secure and reliable information
  transmission,'' {\em Proceedings of IEEE Information Theory Workshop},
  Chengdu, China, pp.~149--153, Oct. 2006.

\bibitem{hamada07expl}
M.~Hamada, ``Constructive conjugate codes for quantum error correction and
  cryptography,'' 2007.
\newblock E-Print arXiv:cs/0703141v2 (cs.IT).

\bibitem{hamada06md}
M.~Hamada, ``Concatenated quantum codes constructible in polynomial time:
  Efficient decoding and error correction,'' {\em IEEE Trans. Information
  Theory}, vol.~54, pp.~5689--5704, Dec. 2008.

\bibitem{csiszar_koerner}
I.~Csisz\'{a}r and J.~K\"{o}rner, {\em Information Theory: Coding Theorems for
  Discrete Memoryless Systems}.
\newblock NY: Academic, 1981.

\bibitem{csiszar98}
I.~Csisz\'{a}r, ``The method of types,'' {\em IEEE Trans. Information Theory},
  vol.~IT-44, pp.~2505--2523, Oct. 1998.

\bibitem{hamada08aqa}
M.~Hamada, ``Algebraic and quantum theoretical approach to coding on wiretap
  channels,'' {\em Proc.\ International Symposium on Communication, Control and
  Signal Processing}, {M}alta, pp.~520--525, Mar. 2008.

\bibitem{DelsartePiret82}
P.~Delsarte and P.~Piret, ``Algebraic construction of {S}hannon codes for
  regular channels,'' {\em IEEE Trans.\ Information Theory}, vol.~28,
  pp.~593--599, July 1982.

\bibitem{hamada05qc}
M.~Hamada, ``Quotient codes and their reliability,'' {\em IPSJ Digital
  Courier}, vol.~1, pp.~450--460, Oct. 2005.
\newblock Available at {\tt
  http://www.jstage.jst.go.jp/article/ipsjdc/1/0/1\_450/\_article}. Also
  appeared in {\em IPSJ Journal},\/ vol.~46, pp.~2428--2438, no.~10, Oct.,
  2005.

\bibitem{cover_th}
T.~M. Cover and J.~A. Thomas, {\em Elements of Information Theory}.
\newblock NY: Wiley, 1991.

\end{thebibliography}
\end{document}